\documentclass[10pt,journal,epsfig,comsoc]{IEEEtran}
\usepackage{graphicx}
\usepackage{subfigure,color}
\usepackage{amssymb}
\usepackage{cite,comment}
\usepackage{amsmath}
\usepackage{bm,comment,url}
\usepackage{algorithm}
\usepackage{algpseudocode}
\makeatletter

\newcommand{\Rmnum}[1]{\expandafter\@slowromancap\romannumeral #1@}

\makeatother

\begin{document}
\title{Aerial-Ground Interference Mitigation for Cellular-Connected UAV}
\author{Weidong Mei and Rui Zhang, \IEEEmembership{Fellow, IEEE}
\thanks{\footnotesize{The authors are with the Department of Electrical and Computer Engineering, National University of Singapore, Singapore 117583 (e-mails: wmei@u.nus.edu, elezhang@nus.edu.sg). W. Mei is also with the NUS Graduate School, National University of Singapore, Singapore 119077.}}}
\markboth{IEEE WIRELESS COMMUNICATIONS}{}
\maketitle

\begin{abstract}
To support large-scale deployment of unmanned aerial vehicles (UAVs) in future, a new wireless communication paradigm, namely, \textbf{\textit{cellular-connected UAV}}, has recently received an upsurge of interests in both academia and industry. Specifically, cellular base stations (BSs) and spectrum are reused to serve UAVs as new aerial user equipments (UEs) for meeting their communication requirements. However, compared to traditional terrestrial UEs, the high altitude of UAVs results in more frequent line-of-sight (LoS) channels with both their associated and non-associated BSs in a much wider area, which causes stronger aerial-ground interference to both UAVs and terrestrial UEs. As such, conventional techniques designed for mitigating the terrestrial interference become ineffective in coping with the new and more severe UAV-terrestrial interference. To tackle this challenge, we propose in this article new interference mitigation solutions for achieving spectral efficient operation of the cellular network with co-existing UAVs and terrestrial UEs. In particular, we exploit the powerful sensing capability of UAVs and inactive BSs in the network for interference mitigation/cancellation. Numerical results are presented to verify the efficacy of the proposed solutions and show their significant spectrum efficiency gains over terrestrial interference mitigation techniques. 
\end{abstract}

\section{Introduction}
The popularity of unmanned aerial vehicles (UAVs) (a.k.a. drones) has increased rapidly over the last decade and is expected to spur even greater usage in future. Recent study has anticipated that the worldwide commercial UAV market will triple in size to 14 billion dollars by 2029. This is mainly attributed to the fact that UAVs have been made more easily accessible to the civilian users than ever before, thanks to their steadily decreasing cost and improved portability. Broader use of UAVs is foreseen in multifarious new applications, such as packet delivery, video surveillance, warehousing and inventory, communication platform, etc. However, the expanding UAV market, in turn, places more stringent requirements on the performance of UAVs' communications\cite{zeng2019accessing}, in order to ensure their ultra-reliable control and non-payload communication (CNPC) as well as support their high-capacity payload communication, such as high-definition video streaming.

However, most of today's commercial UAVs still rely on the simple direct point-to-point communications with their associated ground pilots, which are usually limited to the visual or radio line-of-sight (LoS) range only. As such, both academia and industry have recently looked into a new paradigm, namely, {\it cellular-connected UAV}\cite{zeng2019cellular}, by reusing the existing cellular base stations (BSs) and spectrum to serve UAVs as new aerial user equipments (UEs), in addition to the traditional terrestrial UEs. By leveraging the advanced and well-developed cellular technologies, both CNPC and payload communication requirements for UAVs can be potentially achieved. In fact, preliminary field trials have demonstrated that the current fourth-generation (4G) long-term evolution (LTE) network is already able to meet some basic requirements of UAV-ground communications\cite{lin2019mobile}.

\begin{figure}[!t]
\centering
\includegraphics[width=3.2in]{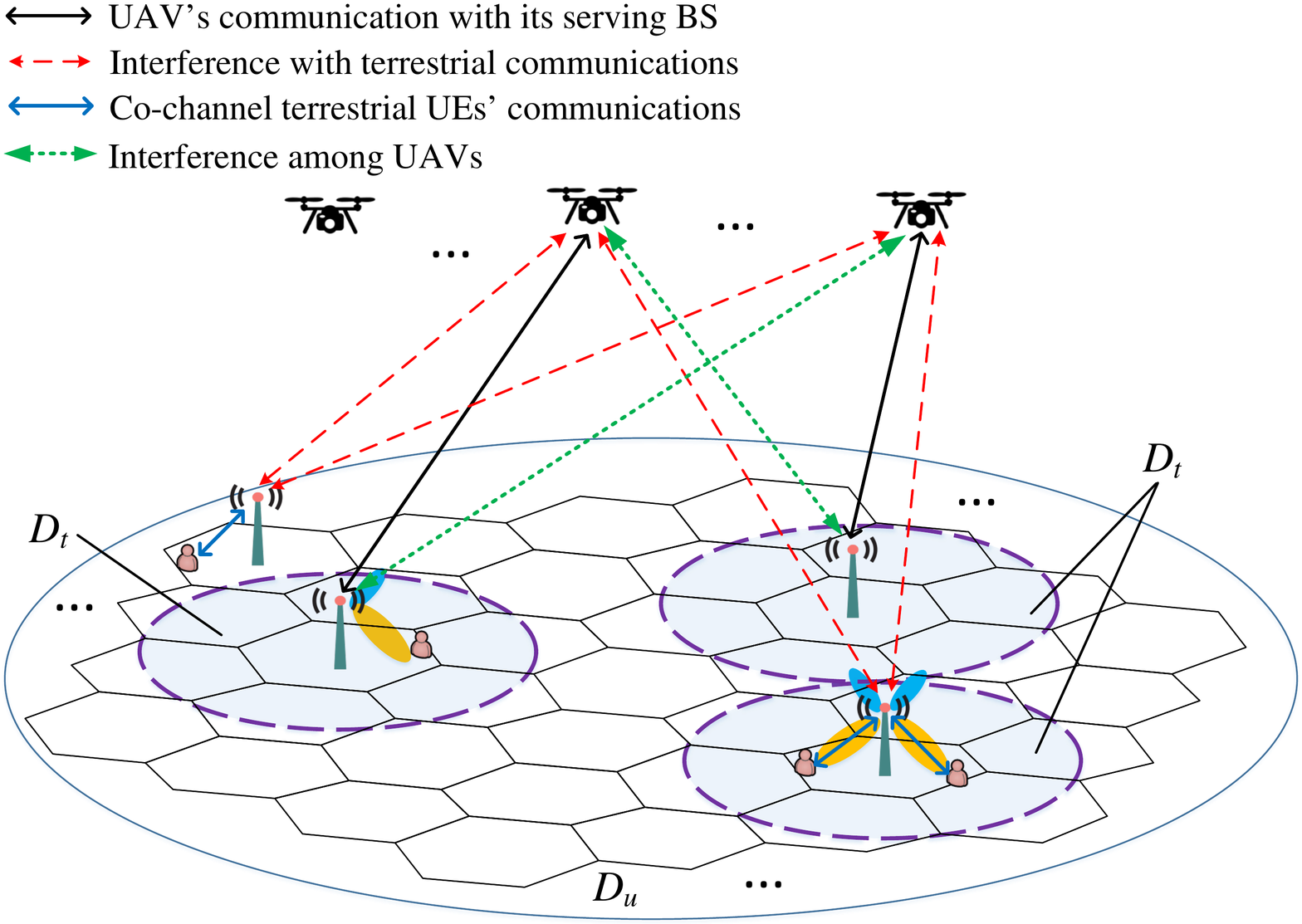}
\DeclareGraphicsExtensions.
\caption{Cellular-connected UAV communications subjected to aerial-ground as well as inter-UAV interferences, where $D_t$ and $D_u$ depict the signal coverage (or required ICIC) region of typical terrestrial UE and UAV, respectively.}\label{ccuc}
\vspace{-12pt}
\end{figure}

Nonetheless, inter-cell interference (ICI) has been identified as the main roadblock to achieving spectral efficient cellular-connected UAV communications\cite{zeng2019cellular,lin2019mobile,fotouhi2019survey}. Due to the high altitude of UAVs, they usually have much more frequent LoS channels with a substantially larger number of ground BSs as compared to terrestrial UEs, regardless of whether they are serving UAVs or terrestrial UEs. As a result, a more severe and complicated interference issue arises as compared to conventional terrestrial ICI, which comprises the aerial-ground interference (between UAVs' and co-channel terrestrial communications) and inter-UAV interference (among co-channel UAV communications), as shown in Fig.\,\ref{ccuc}, in both the uplink (UAV-to-BS) and downlink (BS-to-UAV) communications. How to effectively resolve both types of interference is thus crucial to achieve spectral efficient UAV-cellular integration.

On the other hand, it is worth noting that the presence of LoS links between UAVs and ground BSs also brings advantages for interference mitigation. Firstly, this leads to more reliable communication link between each UAV and its associated BS than terrestrial link, which generally experiences more severe path-loss, shadowing and multi-path fading. Moreover, the LoS channels also make a UAV connectable to more BSs, thus yielding a higher macro-diversity gain in UAV cell association and BS cooperative processing for interference cancellation. For example, as shown in Fig.\,\ref{ccuc}, UAVs can reliably exchange critical control information with many ground BSs for facilitating their ICI coordination (ICIC)\cite{cellular2018mei}. Specifically, each UAV can broadcast the indices of the resource blocks (RBs) assigned to it by its serving BS to all other BSs in $D_u$ by leveraging its reliable uplink control channels with them, so that they will not assign these RBs to their served UAVs and thus avoid the inter-UAV interference in both the uplink and downlink transmissions.

However, compared with the terrestrial interference and inter-UAV interference, the aerial-ground interference is more challenging to mitigate. Note that this interference issue cannot be simply resolved by assigning UAVs with only those RBs that are not used by any terrestrial UEs. This is because the number of RBs available to the UAVs in this case would be highly limited or even zero with a high probability due to the dense frequency reuse for terrestrial UEs in today's cellular network and their current number significantly surpasses that of UAVs. Furthermore, existing terrestrial ICI mitigation techniques have their limitations in dealing with the more severe aerial-ground interference, elaborated as follows.
\begin{itemize}
	\item {\it Dynamic frequency reuse and terrestrial ICIC:} In terrestrial network, adjacent BSs share control information with each other via the high-speed backhaul links (e.g., X2 interface in LTE) to dynamically allocate RBs to their served UEs, so as to avoid the ICI. Due to the significant path-loss over terrestrial channels, such ICIC only needs to involve a few BSs in a local region (see $D_t$ in Fig.\,\ref{ccuc}) to exchange information. However, when this technique (termed as the terrestrial ICIC) is applied at the serving BS of a UAV, a much larger number of co-channel BSs outside $D_t$ (but still in $D_u$) are overlooked, thus causing severe aerial-ground interference between the UAV and co-channel terrestrial communications.
	\item {\it Coordinated multipoint (CoMP) processing:} In CoMP, adjacent BSs virtually form a large distributed antenna system by sharing the UE messages/signals via the backhaul links for jointly precoding/decoding in the downlink/uplink communication, thus enhancing the interference mitigation over the terrestrial ICIC\cite{gesbert2010multi}. While large-scale CoMP can satisfactorily resolve the aerial-ground interference issue by engaging all BSs in $D_u$ to cooperatively serve terrestrial UEs and UAVs simultaneously, it requires involving much more BSs as compared to the conventional CoMP applied in each local region $D_t$ for terrestrial UEs only, thus greatly increasing the processing complexity/delay and signal exchange overhead at the BSs.
	\item {\it Non-orthogonal multiple access (NOMA):} At first glance, it is an appealing approach to apply NOMA to cancel the aerial-ground interference via successive interference cancellation (SIC), since the highly asymmetric (LoS versus non-LoS (NLoS)) channels between UAV and terrestrial UE with the ground BS resemble the near-far scenario desirable in the terrestrial NOMA\cite{ding2017survey} where a near UE to the BS has stronger channel than a far-away UE and thus SIC can be efficiently applied to decode the far/near UE's message before decoding the other UE's message in the downlink/uplink NOMA. However, due to the existence of a large number of co-channel BSs in $D_u$, the UAV's achievable rate in the uplink NOMA would be severely limited by the co-channel BS with the worst channel condition with it (for canceling its signal at the co-channel BS before decoding the terrestrial UE's signal via SIC); whereas in the downlink NOMA, each UAV needs to decode a large number of co-channel terrestrial UEs' messages (before decoding its own message via SIC), which is practically difficult or even infeasible.
	\item {\it Multi-user multi-input multi-output (MU-MIMO):} With MU-MIMO, multiple terrestrial UEs can be served by a multi-antenna BS in the same time-frequency RB via separating their signals in the spatial domain. However, in the current 4G LTE network, MU-MIMO is mainly implemented in the horizontal (azimuth) dimension, while it has very limited capability of spatial interference suppression in the elevation domain\cite{zeng2019cellular}. This is because practical BSs are usually equipped with directional antennas with fixed downtilt. For example, as shown in Fig.\,\ref{ccuc}, the BS can serve two terrestrial UEs with different horizontal angles with it simultaneously using two beams at the same frequency. However, it is unable to separate the signal of each terrestrial UE from that of a high-altitude UAV even with drastically different elevation angles with them.
\end{itemize}
The above shows that the conventional interference mitigation techniques for terrestrial networks are insufficient to deal with the more challenging aerial-ground interference issue, which thus motivates this article to propose new and more effective solutions. In particular, this article focuses on presenting the new solutions that can be implemented in the current cellular network without the need of significantly changing its infrastructure (say, equipping all BSs with fully digital three-dimensional (3D) massive MIMO\cite{kammoun2019elevation}), by exploiting UAV's sensing and idle BSs in the network for ICIC.

\section{Aerial-Ground Interference Mitigation Solutions}
In this section, we propose various solutions to tackle the new UAV-terrestrial interference problem. For the purpose of exposition, we assume that the UAVs in the same region are assigned into orthogonal RBs for communications with their associated BSs and thus there is no inter-UAV interference. As such, we can focus on dealing with the UAV-terrestrial interference in the network.

\subsection{RB Allocation Strategies}
\begin{figure}[!t]
\centering
\includegraphics[width=3.4in]{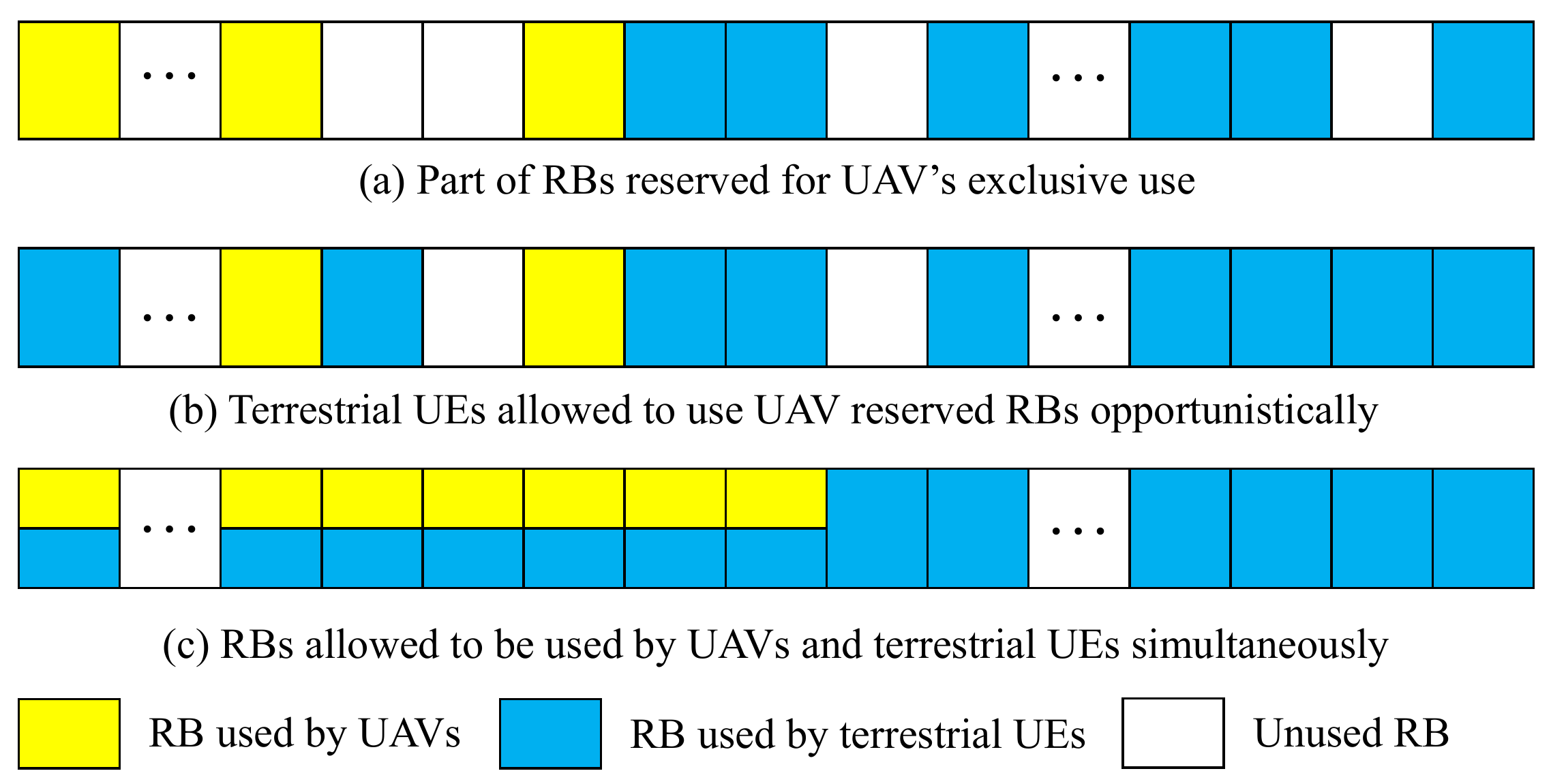}
\DeclareGraphicsExtensions.
\caption{RB allocation strategies for terrestrial UEs and UAVs.}\label{manage}
\vspace{-12pt}
\end{figure}
When the number of UAVs is small or moderate in a given region, a straightforward solution to avoid the aerial-ground interference is by reserving a fraction of the cellular RBs for their exclusive use, regardless of whether they are present or not, as shown in Fig.\,\ref{manage}(a). However, due to the dense frequency reuse in the existing cellular network, this will result in huge spectrum efficiency loss of the terrestrial communications since these reserved RBs cannot be used by terrestrial UEs in all cells. Moreover, the percentage of RBs reserved for UAVs needs to increase with their growing number in the future, thus resulting in more spectrum efficiency loss. An improved solution could be that allowing the terrestrial UEs to access the RBs reserved for UAVs opportunistically\cite{zhao2007survey}, i.e., when they are not present (see Fig.\,\ref{manage}(b)). However, as long as a new UAV requests communication and broadcasts the indices of its assigned RBs to ground BSs, all the existing terrestrial UEs using these RBs would need to immediately evacuate to other available RBs. As such, if all the reserved RBs are taken by UAVs and the terrestrial UE density is high, the remaining RBs would be insufficient for terrestrial UEs, which results in their low rates. On the other hand, if the UAV density is high, the reserved RBs would be insufficient for supporting their communications too. In such cases, the only possible solution would be spectrum sharing between UAVs and terrestrial UEs over the UAV reserved and even non-reserved RBs (see Fig.\,\ref{manage}(c)), provided that their mutual interference can be effectively mitigated. To this end, we propose various solutions in the next. 

\subsection{UAV Sensing-Assisted ICIC}
As shown in Fig.\,\ref{ccuc}, the major limitation of conventional terrestrial ICIC for cellular-connected UAVs lies in that the UAV's serving BS can only exchange RB allocation information with its adjacent BSs in $D_t$, but has no knowledge about the RB allocations in other cells in the much larger region $D_u$. However, such knowledge is difficult to obtain at the UAV's serving BS, unless there is a global information exchange among all the BSs in $D_u$, which is practically costly to implement.

Fortunately, the UAV's LoS-dominant channel with the terrestrial BSs and UEs endows it with a powerful spectrum sensing capability for detecting strong terrestrial signals over a much larger region than its serving BS\footnote{Note that if the channel between the UAV and any co-channel terrestrial BS/UE is NLoS, then its strength is much weaker as compared to the case of LoS and thus is not our main consideration here. Nevertheless, our proposed solutions apply to both LoS and NLoS channels.}. By adopting the various spectrum sensing techniques designed for the cognitive radio (CR) network\cite{zhao2007survey}, each UAV can detect the transmissions of terrestrial BSs/UEs in a wide area of $D_u$, thereby assisting its serving BS in RB allocation to avoid the interference with terrestrial communications. Specifically, in the downlink, the UAV may sense its received interference over the available RBs at its serving BS to help select low-interference RBs for maximizing its achievable rate. Whereas in the uplink, the UAV may sense the terrestrial UEs' transmissions (and co-channel UAVs' transmissions in the case with inter-UAV interference) in each available RB. If the received power in an RB is high, then it is a good indication that there may exist terrestrial UEs transmitting over this RB in the cells outside $D_t$ but in $D_u$. Thus, the transmit power of the UAV in this RB should be kept low or even set to zero for avoiding its uplink interference with co-channel terrestrial uplink transmissions. Note that the above UAV sensing-assisted ICIC requires only local and distributed cooperation between the UAVs and their serving BSs, but significantly enlarges the interference coordination region as compared to the conventional terrestrial ICIC (i.e., $D_u$ versus $D_t$ in Fig.\,\ref{ccuc}). 

Although the UAV sensing-assisted ICIC design can effectively avoid or reduce the aerial-ground interference, it treats interference simply as noise and thus the achievable rate for the UAV/co-channel terrestrial UEs may still be limited in the downlink and uplink communications, respectively. On the other hand, in each of the UAV's assigned RB, there usually exist a large number of ``idle'' BSs adjacent to the UAV's serving BS, which are not serving any terrestrial UEs due to the terrestrial ICIC applied in each local region (see $D_t$ in Fig.\,\ref{ccuc}) and thus can be exploited for mitigating the UAV-terrestrial interference. Motivated by the above, we further propose two new cooperative interference cancellation (CIC) schemes for the UAV uplink and downlink communications, respectively, to more effectively suppress the UAV-terrestrial interference, as shown in Fig.\,\ref{UAV_cic}. For ease of explanation, we assume that all channel powers equal to one in Fig.\,\ref{UAV_cic}. 

\subsection{Cooperative Interference Cancellation (CIC)}
\begin{figure*}[!t]
\centering
\includegraphics[width=5in]{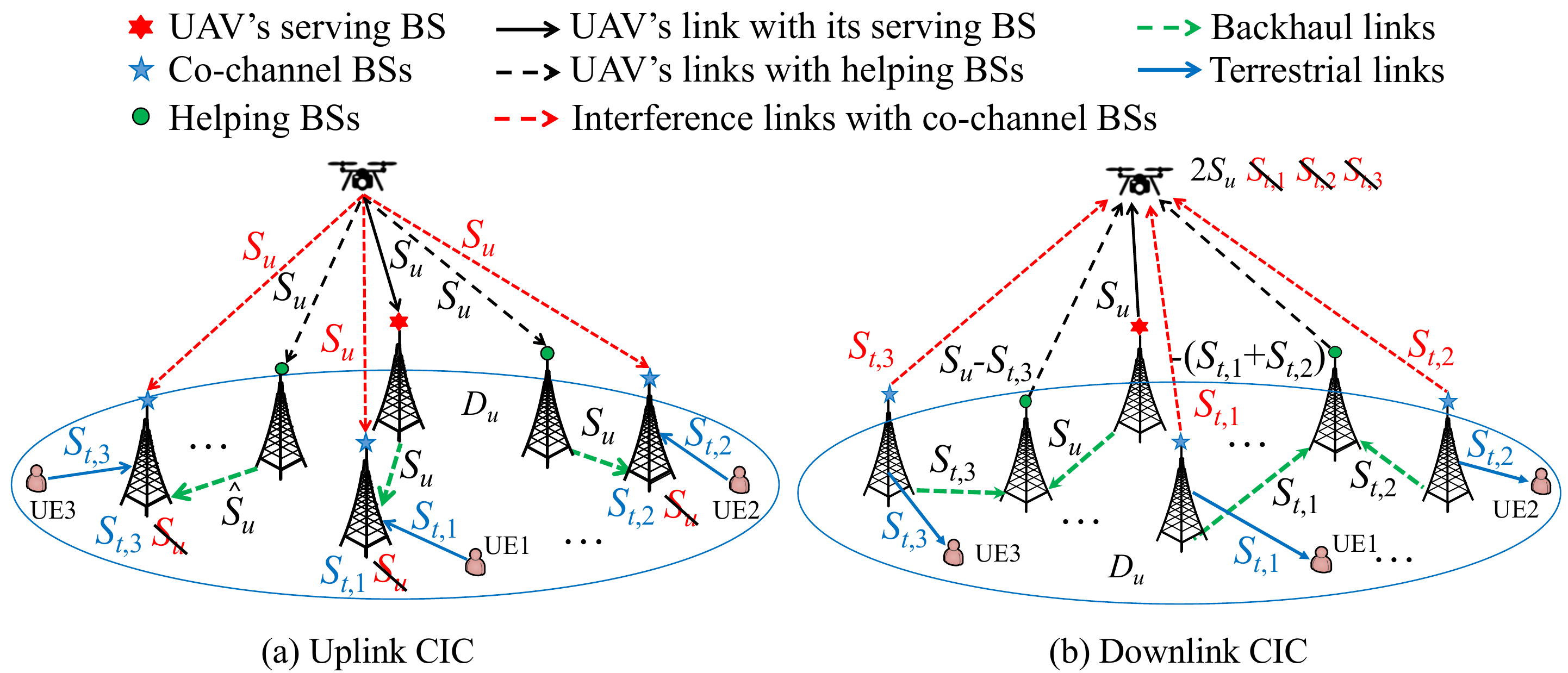}
\DeclareGraphicsExtensions.
\caption{Proposed cooperative interference cancellation (CIC) schemes for cellular-connected UAVs.}\label{UAV_cic}
\vspace{-12pt}
\end{figure*}
{{\it 1)} \bf Uplink CIC:} As shown in Fig.\,\ref{UAV_cic}(a), in addition to the UAV's serving BS, some of the idle BSs in the cellular network can help decode the UAV's signal in the uplink if they have sufficiently good channels with the UAV and are not serving any terrestrial UEs (thus, free of any terrestrial interference) in the UAV's assigned RBs. The chance of finding a large number of such helping BSs is usually very high in practice thanks to the LoS-dominant UAV-BS channels as well as the conventional ICIC for terrestrial UEs which yields many ``idle'' BSs for any used RB in the network. For example, all the outer BSs in each region $D_t$ do not use the same RBs of the BS in the center to serve UEs for mitigating terrestrial ICI. After decoding the UAV's signal, these helping BSs can forward the decoded UAV signal $S_u$ to their adjacent or even farther BSs if they are sharing the same RB of the UAV for serving terrestrial UEs, by exploiting the high-speed and reliable backhaul links (e.g., X2 links) among the BSs. Then these co-channel BSs can cancel $S_u$ before decoding their respective served terrestrial UEs' signals (e.g., $S_{t,1}$ and $S_{t,2}$ shown in Fig.\,\ref{UAV_cic}(a)). As compared to the previous UAV sensing-assisted ICIC, the co-channel BSs can remove the UAV's interference, thus achieving higher rates for the terrestrial UEs in the uplink. In practice, some helping BSs may not have sufficient signal-to-noise ratio (SNR) at their receivers to decode the UAV's signal. In this case, as shown in Fig.\,\ref{UAV_cic}(a), the helping BS can quantize its received UAV signal and forward the quantized signal, denoted by $\hat{S}_u$, to an adjacent co-channel BS, which then cancels the UAV's signal by properly combining the forwarded signal from the helping BS and its own received signal before decoding its served terrestrial UE's signal ($S_{t,3}$). Note that this approach is also applicable in the case of multiple interfering UAVs, where each co-channel BS can perform zero-forcing combining over the forwarded signals to eliminate the inter-UAV interference.

It is worth noting that the uplink CIC is different from the conventional uplink NOMA for terrestrial UEs, as the UAV signal is decoded at the helping BSs (free of terrestrial interference) instead of each of the co-channel BSs via SIC (with the terrestrial UE's signal treated as interference). As a result, the SNRs at these helping BSs are generally significantly higher than the signal-to-interference-plus-noise ratios (SINRs) at the co-channel BSs for decoding the UAV's signal; thus, uplink CIC can greatly improve the UAV's achievable rate over uplink NOMA. In general, uplink CIC can be jointly applied with uplink NOMA to further improve the UAV's rate performance without compromising the terrestrial UEs' rates, termed uplink cooperative NOMA\cite{mei2019uplink}. In addition, uplink CIC is also different from the conventional CoMP decoding, for which all the UAV serving, co-channel and idle BSs need to share their received signals via the backhaul links to jointly decode the signals of the UAV and terrestrial UEs. In contrast, in the proposed uplink CIC, each BS (UAV serving, co-channel, or idle) processes the received signal independently and only adjacent BSs need to share the UAV signals, thus greatly reducing the backhaul signaling overhead and decoding complexity/delay at the cooperating BSs as compared to CoMP applied over all the BSs in $D_u$. 

{{\it 2)} \bf Downlink CIC:} As shown in Fig.\,\ref{UAV_cic}(b), to cancel the terrestrial interference at the UAV receiver in the downlink, each co-channel BS shares its served terrestrial UE's message to its adjacent idle (helping) BSs via the backhaul links. The helping BSs can be flexibly selected based on the rate requirement of co-channel terrestrial UEs. For example, by selecting the BSs outside $D_t$ of each co-channel BS as helping BSs, their downlink transmission will not affect the existing terrestrial communications thanks to the terrestrial ICIC, but can be utilized to enhance the UAV's receive SINR. Specifically, the helping BS can transmit a superposed interference based on the messages forwarded by its adjacent co-channel BSs (e.g., $-(S_{t,1}+S_{t,2})$ shown in Fig.\,\ref{UAV_cic}(b)), such that it is destructively combined with the terrestrial interference signals from these co-channel BSs (i.e., $S_{t,1}$ and $S_{t,2}$) at the UAV's receiver to get them cancelled. As a result, the UAV's achievable rate can be significantly improved in the downlink compared to the previous UAV sensing-assisted ICIC. It is worth noting that in practice, due to the limited power at the helping BS, its transmitted interference may not be sufficiently strong at the UAV's receiver to cancel the terrestrial interference completely; thus, the transmit powers allocated to each terrestrial signal component at the helping BS need to be optimized to maximize the UAV's receive SINR\cite{cooperative2019mei}. To further enhance the UAV's receive SINR, the UAV's message $S_u$ can also be shared to the adjacent helping BSs via the backhaul links and then cooperatively transmitted, such that they are constructively combined with the signal from the UAV's serving BS to maximize the desired signal power at the UAV's receiver. For example, as shown in Fig.\,\ref{UAV_cic}(b), the UAV's serving BS shares $S_u$ to one adjacent helping BS, which then transmits the superposition of $S_u$ and the terrestrial UE's signal $S_{t,3}$ forwarded by another co-channel BS; as a result, at the UAV's receiver, the desired UAV signal becomes $2S_u$ while the terrestrial interference $S_{t,3}$ is completely cancelled. Again, due to the practically limited transmit power at the helping BS, there is an interesting trade-off between allocating the transmit powers over the desired signal and the interference (i.e., $S_u$ versus $S_{t,3}$) to balance between maximizing the UAV signal power and minimizing the terrestrial interference at the UAV's receiver\cite{cooperative2019mei}. Finally, this downlink CIC is also applicable if there exist multiple UAVs in the same RB, where the transmission of $S_u$ at the helping BS achieves a dual objective of enhancing the UAV's desired signal and cancelling terrestrial interference from its aided BS to other co-channel UAVs.

Notice that the above downlink CIC is also different from the conventional downlink NOMA, as the UAV receiver can directly decode its message without performing SIC. Nonetheless, downlink CIC may be used jointly with downlink NOMA, where some helping BSs can cooperatively transmit the terrestrial interference to make it constructively combined with that from the co-channel BS to maximize the interference signal power at the UAV's receiver for decoding it prior to decoding the UAV's signal via SIC. This thus provides another interesting objective for the signal power allocations at the helping BSs. Moreover, similar to the uplink, downlink CIC is different from the conventional CoMP precoding, which requires all BSs (UAV serving, co-channel, and idle) to jointly transmit the UAV's and all terrestrial UEs' signals, thus requiring substantially higher implementation complexity than our proposed downlink CIC. In particular, the power allocation at each helping BS in the downlink CIC can be optimized in a distributed manner requiring only local information exchange\cite{cooperative2019mei}. Thus, both uplink and downlink CIC are scalable with increasing number of terrestrial UEs and UAVs.

\subsection{Exploiting UAV Beamforming}
Exploiting beamforming at the UAV can help enhance the performance of the above proposed interference mitigation solutions. For example, in the UAV sensing-assisted ICIC, the UAV can leverage multiple antennas to learn more refined and useful information about the sensed terrestrial transmissions\cite{zhang2010cognitive}. Specifically, multi-antenna sensing can detect the directions of strong terrestrial transmissions and help the UAV in nulling the interference from/to them via beamforming in its downlink/uplink communication. Besides, in the uplink CIC, in order for all helping BSs to decode the UAV's signal, the design of UAV transmit beamforming is to maximize the UAV's signal SNRs at these helping BSs\cite{liu2018multi}, which resembles that of multicast beamforming for common information transmission to all the helping BSs. It is worth noting that this approach is more effective than the conventional interference nulling at the UAV's transmitter due to the large number of co-channel BSs but only a limited number of UAV antennas\cite{liu2018multi}. Finally, in the downlink CIC, the UAV can perform spatial filtering over its received signals at different antennas to enhance its desired signal and/or null the residual co-channel interference to maximize its SINR. Alternatively, spatial filtering can be used to enhance the interference if its strength is sufficiently strong, so as to facilitate the UAV's SIC, if downlink NOMA and downlink CIC are jointly applied.

\subsection{Interference-Aware UAV Trajectory Design}
\begin{figure}[!t]
\centering
\includegraphics[width=3.4in]{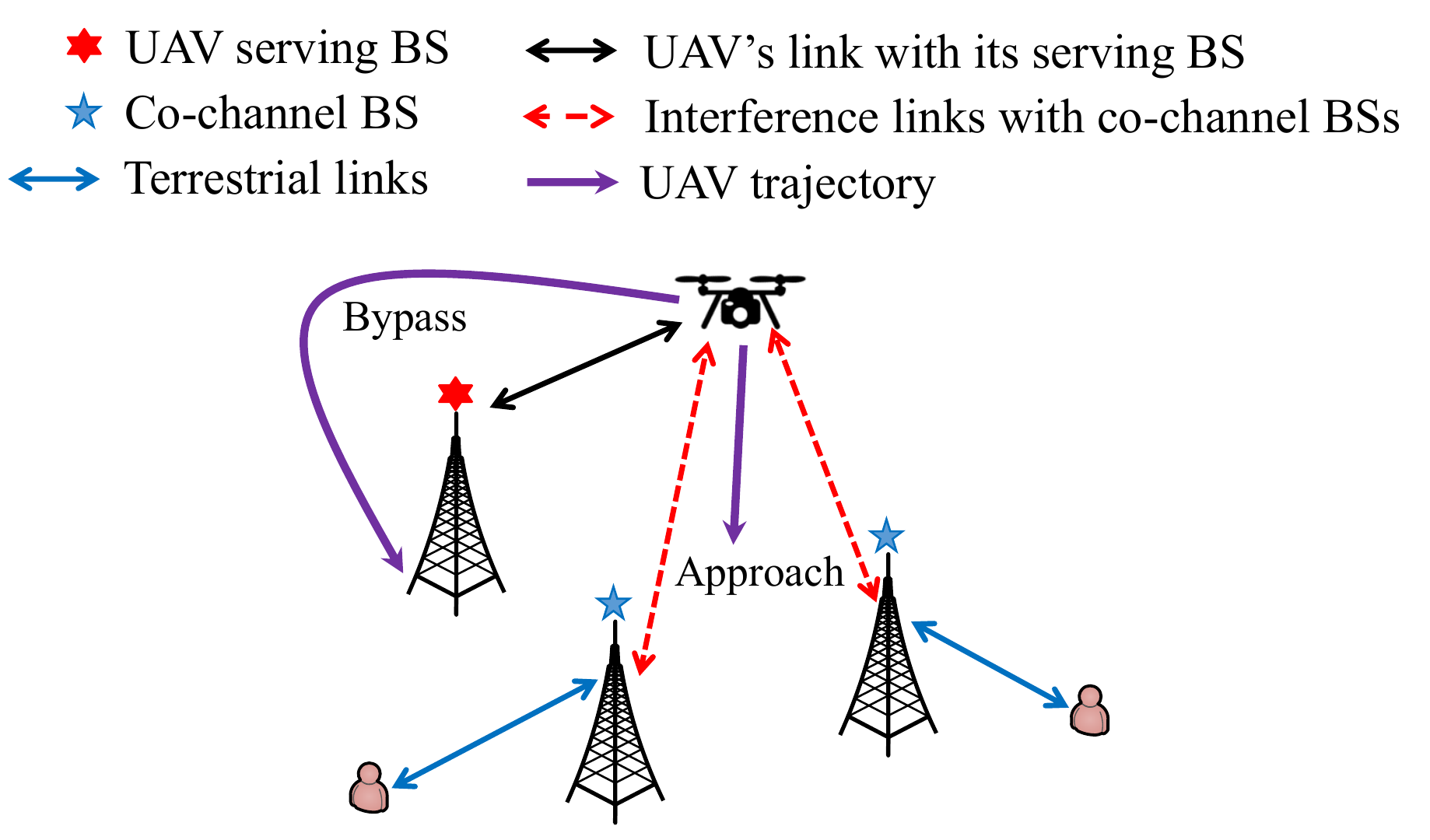}
\DeclareGraphicsExtensions.
\caption{Interference mitigation by exploiting UAV trajectory design.}\label{traj}
\vspace{-12pt}
\end{figure}
Thanks to the high and controllable 3D mobility of UAVs, UAV trajectory design emerges as a new means for mitigating the aerial-ground and inter-UAV interference in both the uplink and downlink communications\cite{huang2019cognitive}. As shown in Fig.\,\ref{traj}, the UAV trajectory can be flexibly designed to either bypass or approach the co-channel BS in its fly direction, so as to avoid its interference with this BS or even enhance it for facilitating the SIC at the UAV/co-channel BS in the downlink/uplink for cancelling the corresponding aerial-ground interference. Moreover, UAV trajectory can be jointly designed with the previously proposed techniques such as UAV sensing, beamforming and/or BSs' CIC to further enhance the UAV-terrestrial interference mitigation performance.

To summarize, among the proposed interference mitigation solutions, UAV sensing-assisted ICIC requires the lowest overhead, as it requires only local cooperation between each UAV and its serving BS. CIC can further improve the performance by engaging more BSs in cooperation, while incurring higher signaling and channel estimation overhead. Finally, UAV beamforming and/or trajectory design can be combined with the above solutions for performance improvement but also incur more overhead.

\section{Simulation Results and Discussion}
In this section, numerical results are provided to demonstrate the efficacy of the proposed interference mitigation solutions. The simulation settings are as follows. The number of UAVs is set to 8. We consider a chunk of the cellular spectrum comprising 10 RBs that may or may not be reserved for the UAVs. Each UAV associates with the BS that has the smallest path-loss with it and is assigned with one RB for its downlink or uplink communication. The bandwidth of each RB is 180 kHz. The carrier frequency is 2 GHz, and the noise power spectrum density at the UAV/BS receiver is $-$164 dBm/Hz. For the terrestrial channels, the path-loss and shadowing are modeled based on the urban macro (UMa) scenario in the 3GPP technical specification\cite{3GPP36777}, while the small-scale fading is modeled as Rayleigh fading. The cell radius is 800 m, and the height of BSs is set to be 25 m. The altitude of the UAV is fixed as 200 m. The BS antenna pattern is assumed to be omnidirectional in the horizontal plane but directional in the vertical plane with 10-degree downtilt angle. We consider three tiers of cells to cover $D_u$ in Fig.\,\ref{ccuc}, and it can be shown that the total number of cells required is 37. The UAVs' and terrestrial UEs' horizontal locations are both randomly generated in the 37 cells. Following the probabilistic LoS channel model in the UMa scenario in \cite{3GPP36777}, all channels between UAVs and terrestrial BSs/UEs are dominated by LoS propagation at the altitude of 200 m. Each UAV or terrestrial UE is assumed to be equipped with a single omnidirectional antenna\cite{3GPP36777}. The transmit power for each BS or UE is set to be 20 dBm.

\begin{figure*}[!t]
\centering
\subfigure[]{\includegraphics[width=0.332\textwidth]{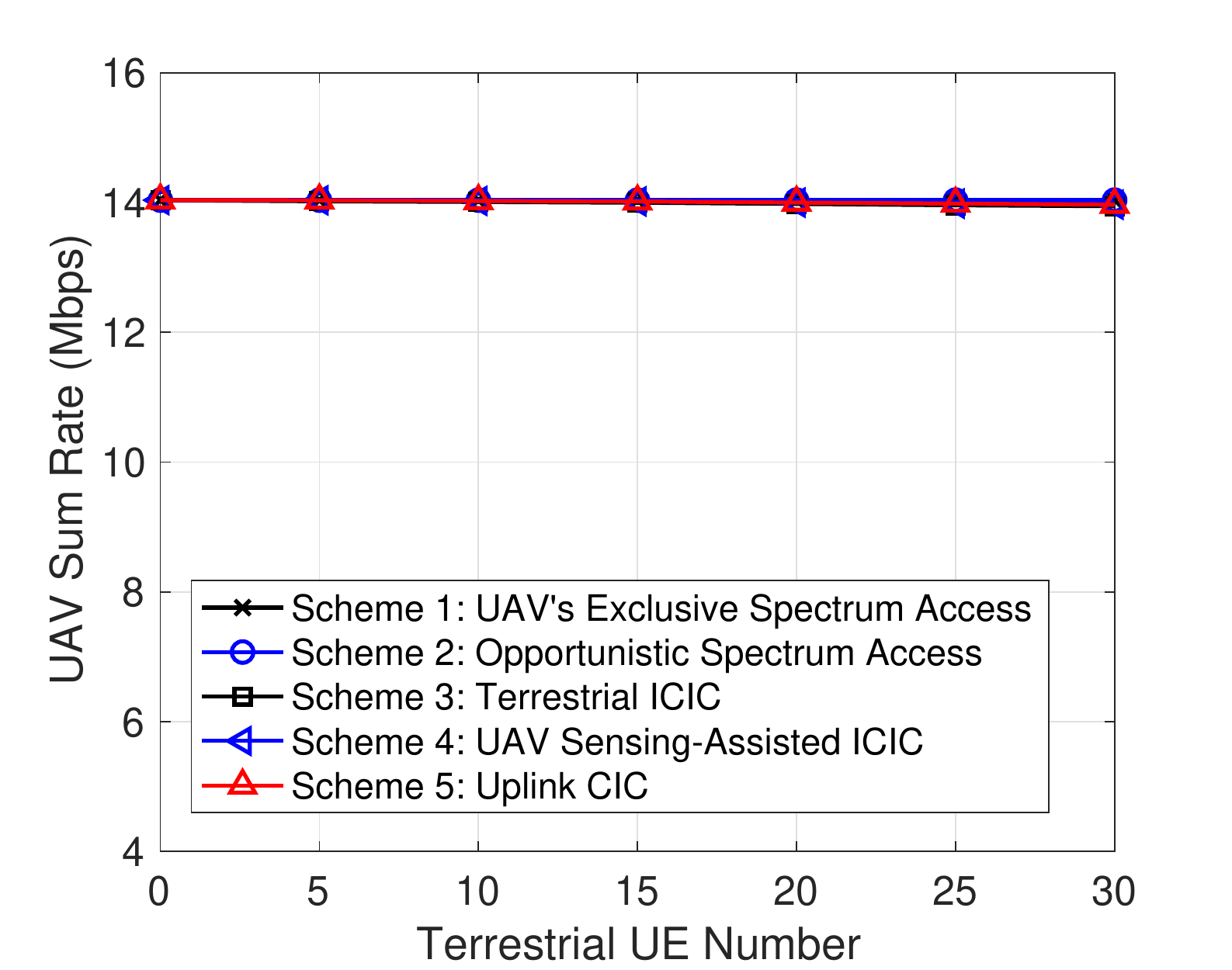}}\!
\subfigure[]{\includegraphics[width=0.332\textwidth]{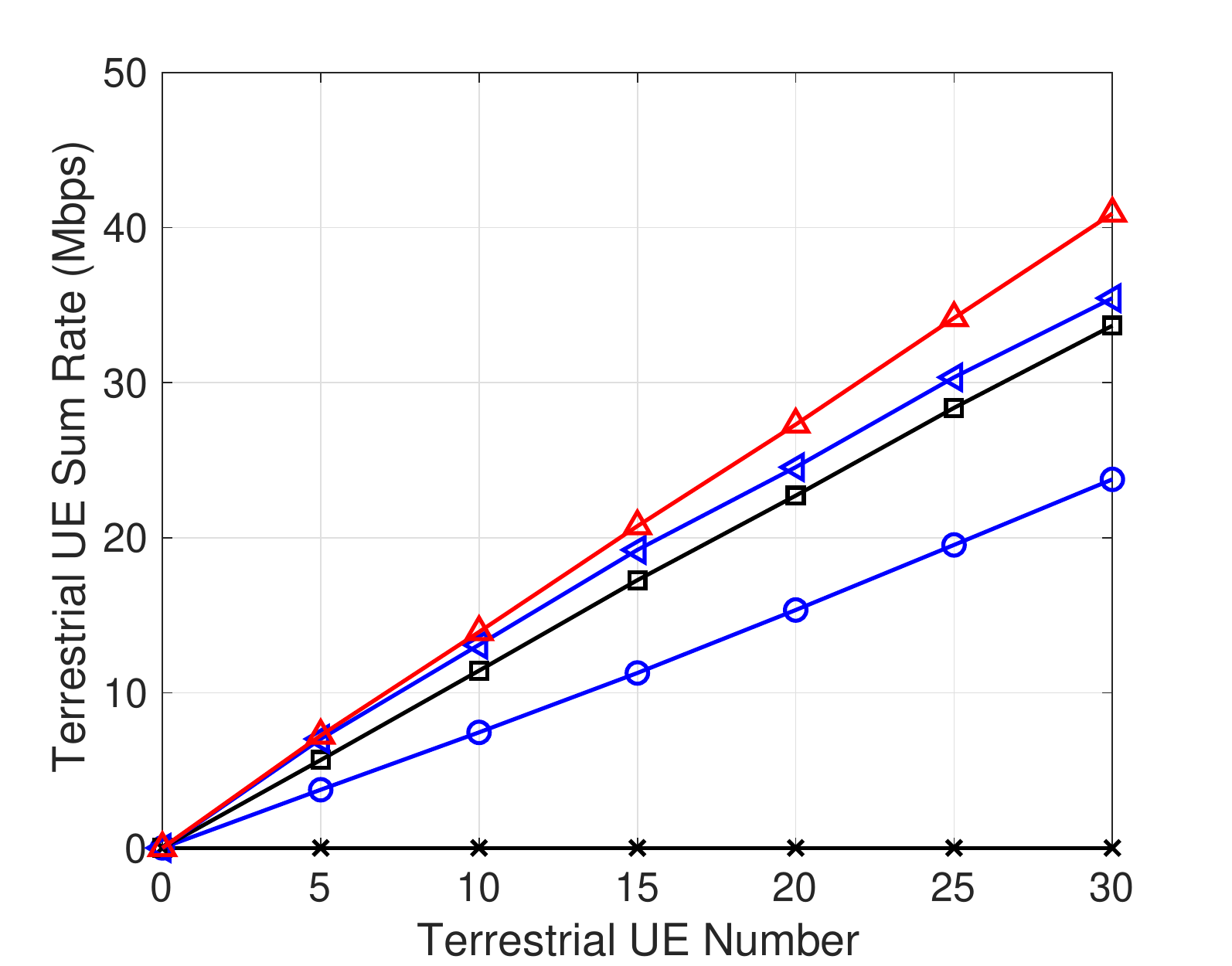}}\!
\subfigure[]{\includegraphics[width=0.332\textwidth]{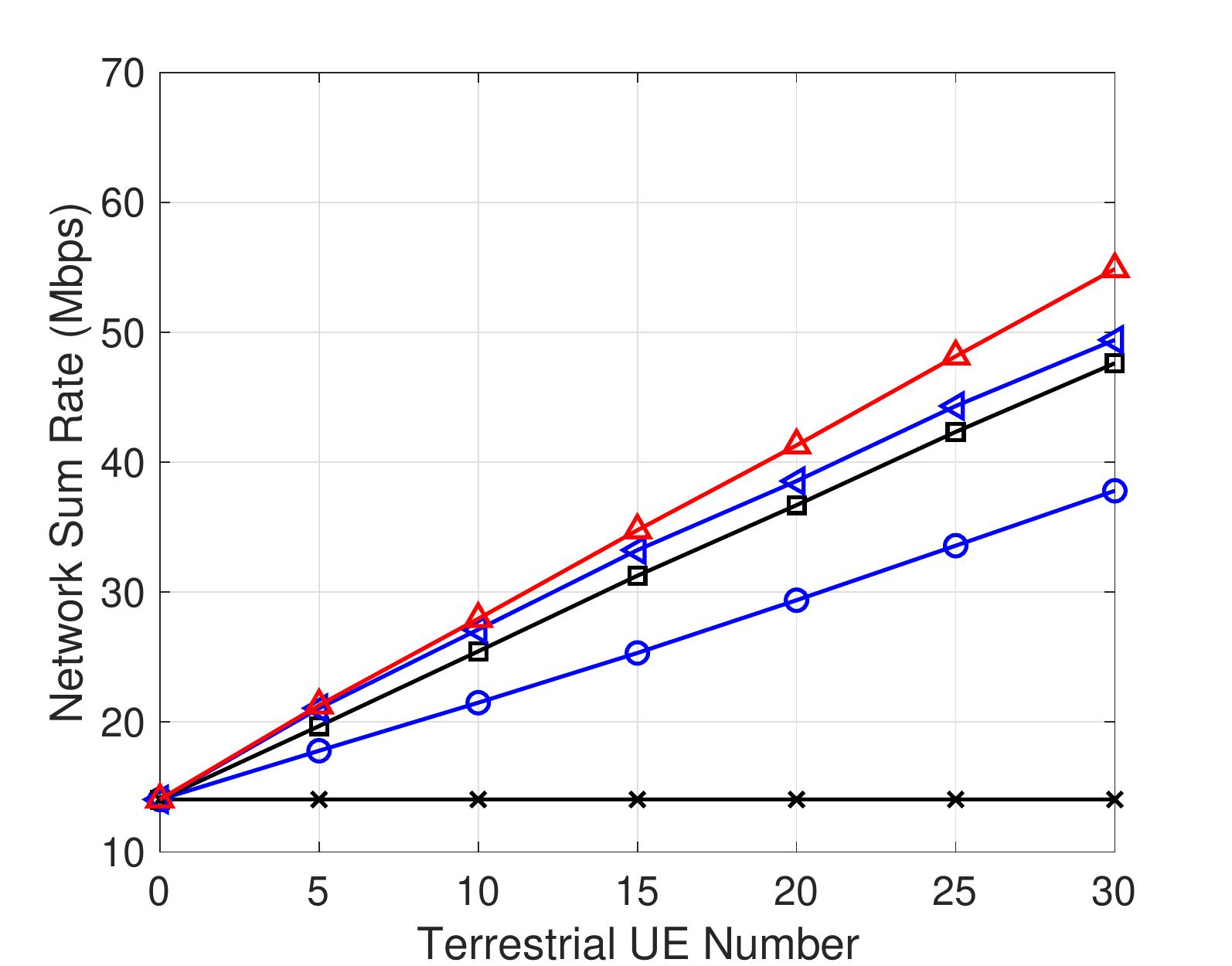}}\!
\caption{Spectral efficiency comparison in the uplink: (a) UAV sum-rate; (b) terrestrial UE sum-rate; (c) network sum-rate.}\label{DensUp}\vspace{6pt}
\centering
\subfigure[]{\includegraphics[width=0.333\textwidth]{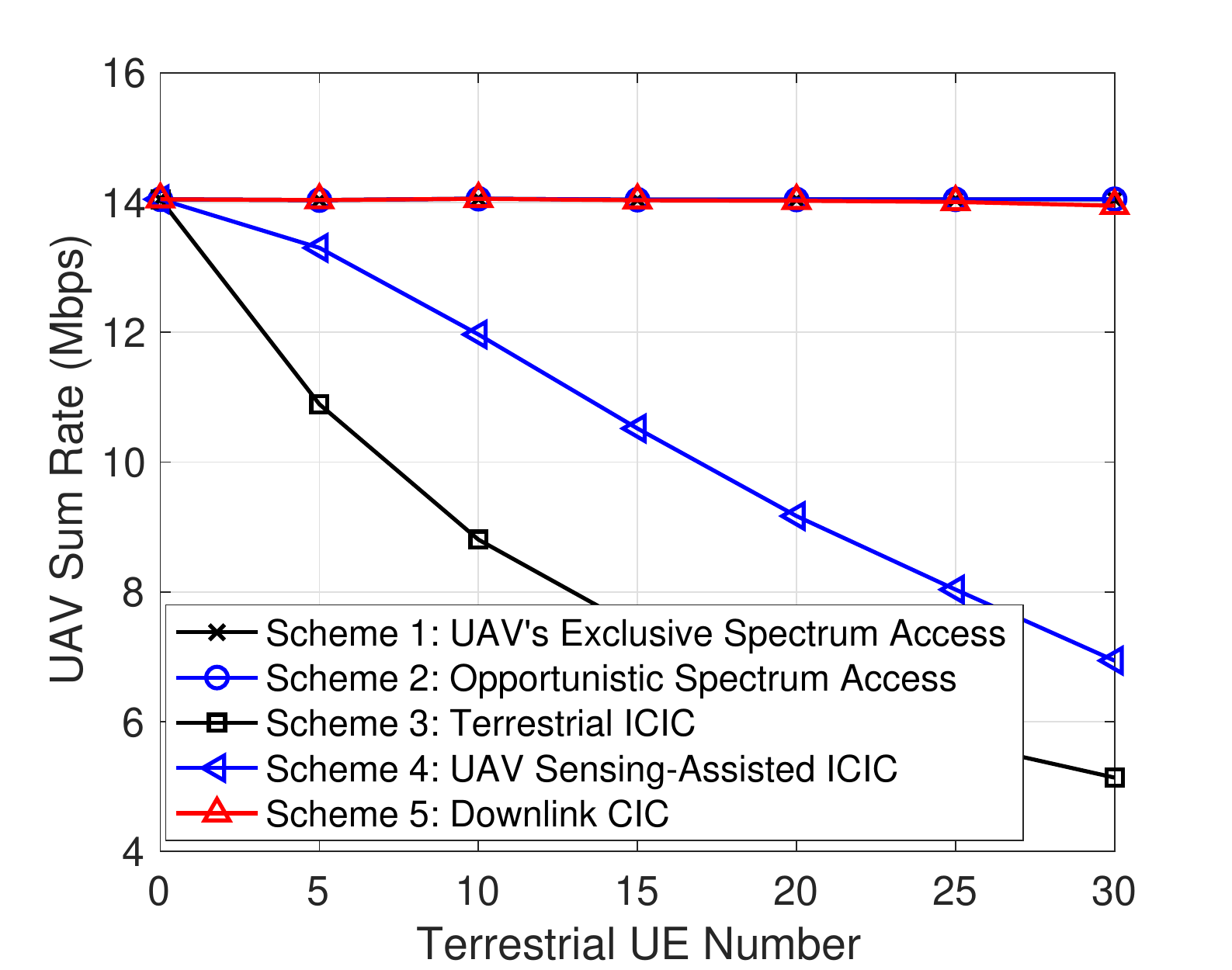}}\!
\subfigure[]{\includegraphics[width=0.332\textwidth]{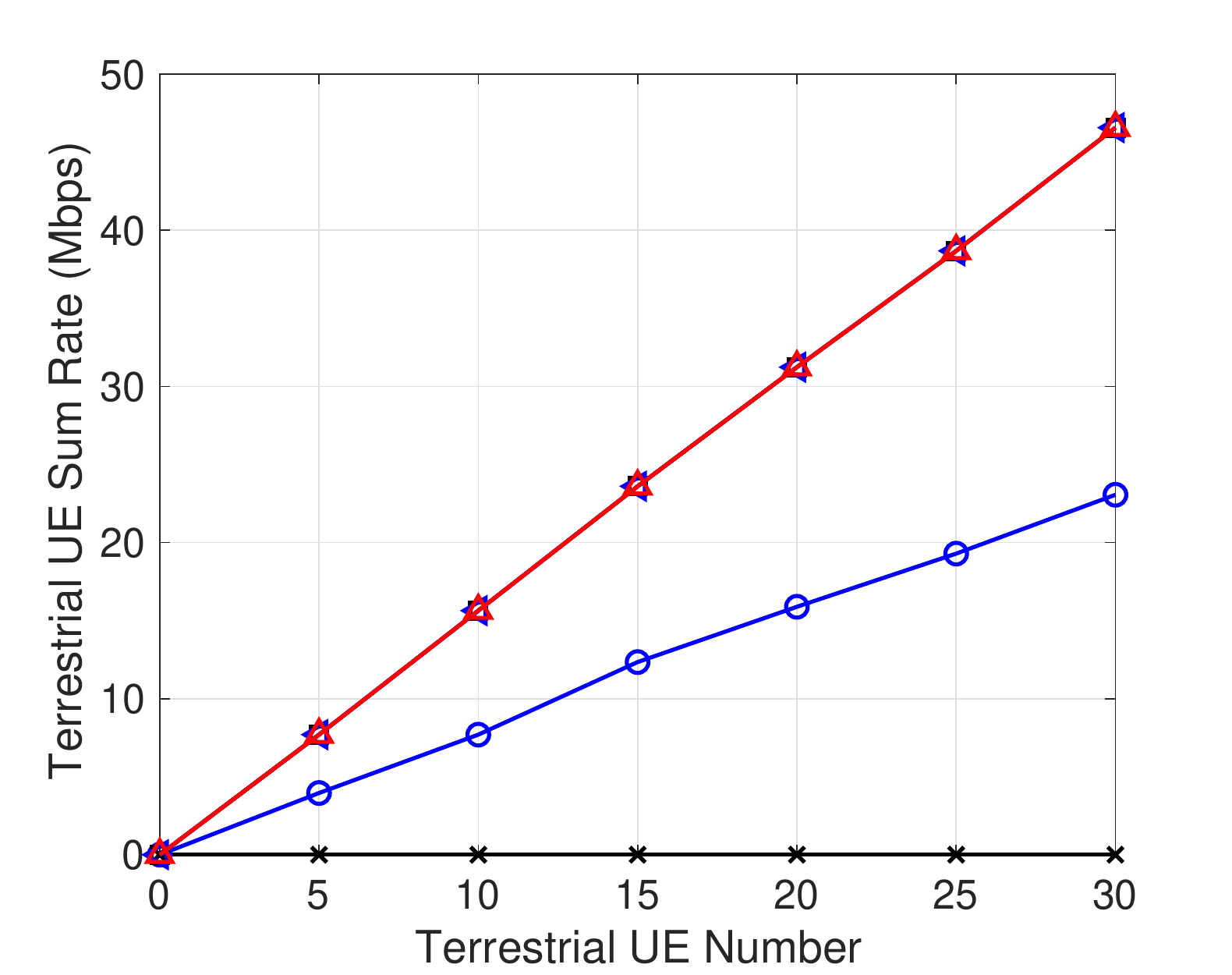}}\!
\subfigure[]{\includegraphics[width=0.332\textwidth]{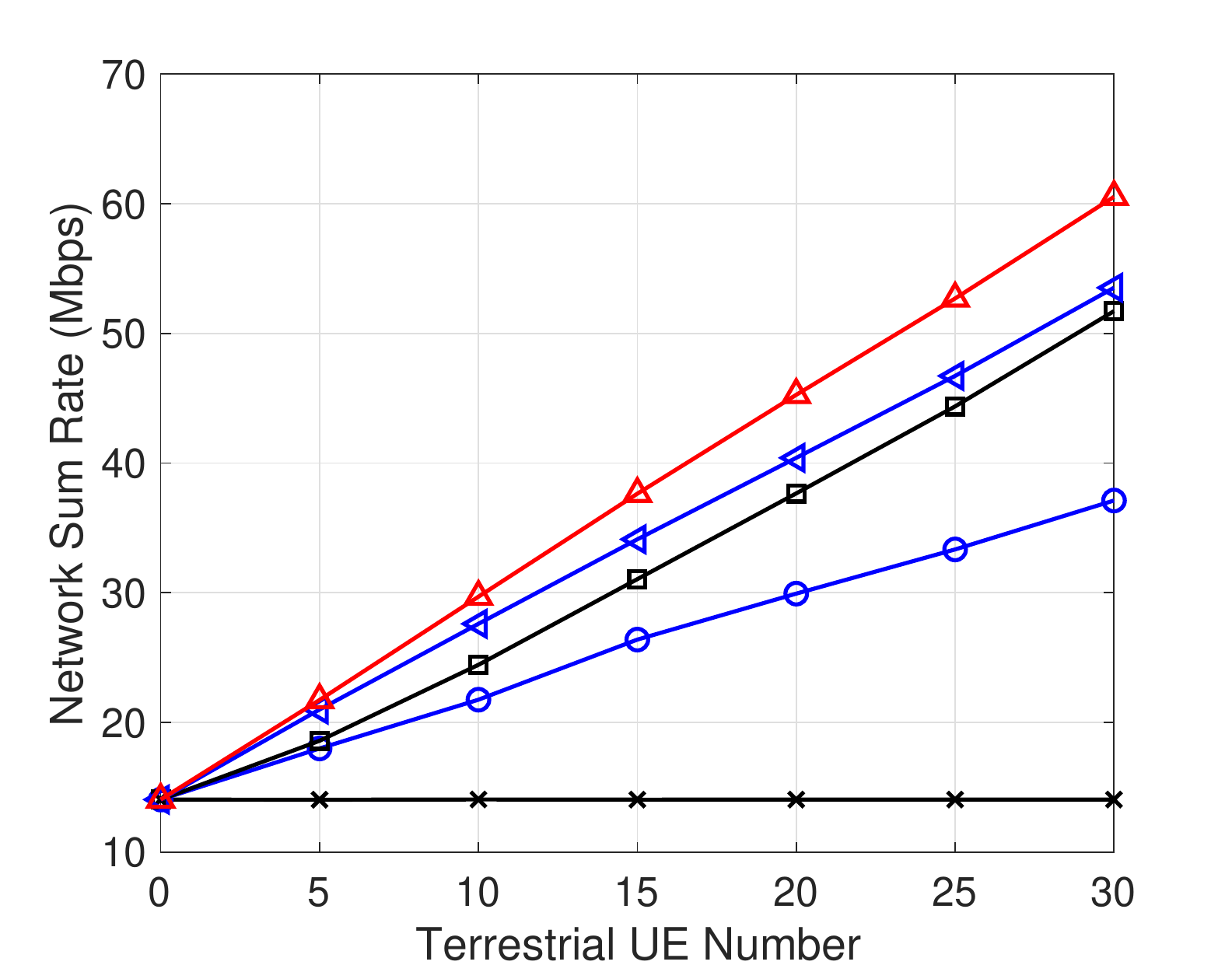}}\!
\caption{Spectrum efficiency comparison in the downlink: (a) UAV sum-rate; (b) terrestrial UE sum-rate; (c) network sum-rate.}\label{DensDown}\vspace{-12pt}
\end{figure*}
In Figs.\,\ref{DensUp} and \ref{DensDown}, we show the spectrum efficiency performance in terms of the UAVs' sum-rate, terrestrial UEs' sum-rate, as well as the network sum-rate over the 10 RBs in the uplink and downlink, respectively, with different numbers of terrestrial UEs. The following five schemes are considered for both uplink and downlink communications, namely, {\it 1)} UAV's exclusive spectrum access for the reserved 10 RBs (see Fig.\,\ref{manage}(a)), {\it 2)} opportunistic spectrum access where terrestrial UEs can use any of the 10 RBs if it is unused by all UAVs (see Fig.\,\ref{manage}(b)), {\it 3)} terrestrial ICIC where UAVs are treated the same as terrestrial UEs to access any of the 10 RBs, {\it 4)} UAV sensing-assisted ICIC, and {\it 5)} uplink/downlink CIC (termed as scheme 1 to 5 in the sequel for brevity). In schemes 1 and 2, each UAV broadcasts the index of its assigned RB to all BSs in $D_u$ to avoid inter-UAV interference. In the terrestrial ICIC (scheme 3), each UAV's serving BS also checks the RB allocation with its nearest neighboring BSs, and then randomly assigns its served UAV with an available RB (if any) that has not been used by them, as well as itself and other UAVs. In contrast, in the UAV sensing-assisted ICIC (scheme 4), each UAV further performs spectrum sensing over all available RBs at its serving BS and is assigned with the RB with the lowest sensed power among them. While in the uplink/downlink CIC (scheme 5), after performing UAV spectrum sensing, each UAV's serving BS/co-channel BS can forward the decoded UAV's/served terrestrial UE's message to its nearest and second-nearest co-channel/idle BSs for interference cancellation. 

From Fig.\,\ref{DensUp}(a), it is observed that with schemes 1 and 2, the UAVs' sum-rate in the uplink is unaffected by the terrestrial UE number, since there is no spectrum sharing between UAVs and terrestrial UEs in these two schemes. However, as observed from Fig.\,\ref{DensUp}(b), this severely limits the terrestrial UEs' sum-rate. In particular, in scheme 1, as the terrestrial UEs are not allowed to use any reserved RB for the UAVs, their sum-rate is zero. In addition, it is observed from Fig.\,\ref{DensUp}(a) that for schemes 3-5 that allow UAV-terrestrial spectrum sharing, the UAVs' sum-rate in the uplink is almost unchanged with increasing the terrestrial UE number. This is because the RB allocation in terrestrial ICIC has been sufficient to mitigate the terrestrial interference at the UAVs' serving BSs. As observed from Fig.\,\ref{DensUp}(b), scheme 4 yields a higher terrestrial UEs' sum-rate than scheme 3, due to the enlarged ICIC region by UAV spectrum sensing; while scheme 5 leads to an even better performance than scheme 4, thanks to the uplink CIC. Finally, it is observed from Fig.\,\ref{DensUp}(c) that by allowing UAV-terrestrial spectrum sharing, the network sum-rate can be greatly improved as compared to that in scheme 1 or 2, since terrestrial UEs are allowed to use the same RBs as the UAVs and there are much more terrestrial UEs than UAVs in the considered network. The proposed schemes 4 and 5 are also observed to outperform scheme 3 thanks to their more effective interference mitigation. The above observations for the uplink are similarly made for the downlink as shown in Fig.\,\ref{DensDown}. In particular, it is observed from Fig.\,\ref{DensDown}(a) that for schemes 3-5 with UAV/terrestrial UE spectrum sharing, the UAVs' sum-rate decreases with the increasing terrestrial UE number, as this also increases the co-channel BS number and thus the terrestrial interference at the UAVs' receivers. Nonetheless, scheme 5 yields considerably better performance than the other two schemes thanks to the downlink CIC. 

It follows from the above results that allowing spectrum sharing between UAVs and terrestrial UEs is an effective means to ensure their coexistence. To address the resultant interference issue, interference mitigation region should be effectively enlarged as compared to that in terrestrial ICIC. To this end, UAVs may play a proactive role by leveraging their strong spectrum sensing capability. Finally, idle BSs can be exploited as helping BSs to tackle the interference issue.

\section{Conclusions}
In this article, we introduce the new and severe aerial-ground interference issue arising from the cellular-connected UAV communications due to the LoS-dominant UAV-ground channels. In view of the limitations of existing terrestrial interference mitigation techniques, we propose new and effective solutions based on the current cellular infrastructure. Simulation results are provided to demonstrate the spectrum efficiency gains by the proposed solutions as compared to conventional techniques for terrestrial interference mitigation. It is hoped that this article will serve as a stepping-stone for the future investigation of spectral efficient interference mitigation techniques for cellular-connected UAV communications.

\bibliography{UAV_Mag}
\bibliographystyle{IEEEtran}

\end{document}